\documentclass{kluwer}
\usepackage[dvips]{graphicx}
\newdisplay{guess}{Conjecture}
\begin{document}
\begin{article}

\begin{opening}
\title{COSMIC DUST IN THE 21ST CENTURY}
\author{J. Mayo \surname{Greenberg}}
\author{Chuanjian \surname{Shen}}
\runningauthor{Greenberg \& Shen}
\runningtitle{Cosmic Dust in the 21st Century}
\institute{The Raymond and Beverly Sackler Laboratory for Astrophysics at Leiden Observatory,
           P.O. Box 9513, 2300 RA Leiden, The Netherlands}

\begin{abstract}
     The past century of interstellar dust has brought us from first ignoring it to finding that it plays an
important role in the evolution of galaxies. Current observational results in our galaxy provide a complex
physical and chemical evolutionary picture of interstellar dust starting with the formation of small
refractory particles in stellar atmospheres to their modification in diffuse and molecular clouds and
ultimately to their contribution to star forming regions. Observations of the properties of dust in very
young galaxies will be an important probe of the rates of star formation in terms of the production and
destruction of dust grains. Future observations of dust at high spectral and spatial resolution will provide
detailed information on processes in collapsing clouds up to star formation. Space missions to comets in
the next century will first study them in situ but ultimately will bring back pristine nucleus material which
will contain the end product of the collapsing protosolar molecular cloud at the time of planet formation.
If one of the current theories of the origin of life from comets is correct, laboratory studies of comet dust
grains immersed in water may give direct indications of prebiotic chemical evolution.
\end{abstract}
\end{opening}
             
\section{Introduction}
                                
     When this title was suggested (not by me) I accepted with great trepidation the chance to make
predictions on what we might find in the coming years about dust. I was very mindful of the possibility
of making too timid predictions as was done by a famous physicist at the turn of the last century who
said that the only new thing would be improvements in accuracy of measurement. Within less than 5
years we had the quantum photoelectric effect, the special theory of relativity, and within 12 years the
Bohr quantum theory of the atom. In any case all I will try to do is to make projections based on what we
now know and to make guesses on how rapidly the observational and theoretical techniques and
programs will evolve in the next 20 years. I hope I can keep a sense of perspective. I think it will be
useful first to briefly summarize some of the high points of what we have learned about dust in the past
century and how we have learned it. I will start with a brief historical review and follow up with a
selection of the present state of the art, with a personal assessment which leads to some projections
which appear to me to be important. I felt that a limited but directed summary would make it much
easier to make specific predictions.

     Interstellar dust has become one of the subjects in the forefront of astrophysics. Dust has to do
with star formation. Infrared observations are a probe of regions of star formation. The interstellar
chemistry problem concerns the chemical reactions between dust and atoms and molecules in space;
dust seems to play an active role in those reactions. The evidence is that dust is one of the basic
ingredients in comets. There is growing evidence from cometary observations that comets are a storage
place for the chemical evolution which takes place in interstellar space. The complex chemistry and
molecule evolution leading to what is now seen in comets may be the necessary precursor to life on
earth and there is reason to believe from the evidence available that the oceans on earth were made of
comets bringing interstellar ice to our young planet.

\section{Early History}

     The subject of dark nebulae and what makes them dark have a very patchy history. The earliest
relevant ideas go back 100 years. Looking at the sky in the direction of Sagittarius it is clear that there
are tremendously dark lanes especially in the region toward the galactic center. This observation inspired
Herschel in 1884 to say that there was undoubtedly a hole in the sky. Barnard started to take pictures
and reported vast and wonderful cloud forms with their remarkable structure, lanes, holes and black
gaps. Agnes Clerke (1903), in an astrophysics text she authored around the turn of the century, 
stated ``The fact is a general one, that in all the forest of the universe there 
are glades and clearings. How they
come to be thus diversified we cannot pretend to say; but we can see that the peculiarity is structural---
that it is an outcome of the fundamental laws governing the distribution of cosmic matter. Hence the
futility of trying to explain its origin, as a consequence, for instance, of the stoppage of light by the
interposition of obscure bodies , or aggregations of bodies, invisibly thronging space.''

     Curtis and Shapley held a famous debate around 1920 (Shapley \& Curtis, 1921) 
over whether what is seen as
the dark lanes in the Milky Way is caused by obscuring material. Curtis said the dark lanes observed in
our galaxy were obscuring material, while Shapley said he found no evidence of obscuring material in
his observations of globular clusters. Later observers became aware that Shapley's argument was
irrelevant because the globular clusters are out of the plane of the galaxy. The obscuring dust was
confined to the so called ``plane of avoidance'' which is the galactic plane. A patchy history indeed.

     Back in 1847 Struve had already found that the number of stars per unit volume seems to
diminish in all directions receding from the sun. This could be explained either if the sun was at the
center of a true stellar condensation, or if the effect was only an apparent one due to absorption (which
may have been understood to include light scattering).

     Well before the Curtis-Shapley debate, Kapteyn (1904) had found a roughly spherical
distribution of stars around the sun, but he did not take that seriously. He assumed a constant stellar
density and then used the observed density to arrive at a value for the extinction (absorption) of light A
= 1.6 mag kpc$^{-1}$, which differs little from current values. In  1929 Schal\'{e}n examined the question of
stellar densities as a function of distance. He did a very detailed study of B and A stars, including those
in Cygnus, Cepheus, Cassiopeia, and Auriga. He obtained rather different values of the absorption
coefficient, particularly in Cygnus and Auriga where there are large dark patches. So obviously the
absorption is more in some regions and less in others.

      It was not until the work of Trumpler in 1930 that the first evidence for interstellar reddening
was found. Trumpler based this on his study of open clusters in which he compared the luminosities and
distances of open clusters with the distances obtained by assuming that all their diameters were the
same. By observing the luminosities and knowing the spectral distribution of stars he was able to find a
color excess (between photographic and visual) with increasing distance, and produce a reddening
curve. His observations indicated reddening even where he saw no clouds. In his 1957 book, Dufay
questioned whether interstellar space outside dark clouds and nebulae should be considered perfectly
transparent. By 1939 (Stebbins, Huffer \& Whitford 1939) such observations were already indicating a
$\lambda ^{-1}$ ``law'' of reddening. But what caused it? Since hyperbolic meteors were thought to exist, first
attempts were made to tie the interstellar dust to the meteors. This was the conjecture of Schal\'{e}n (1929) and
Greenstein (1934). Another possible influencing factor might have been the fact that it was easier to
compute the scattering by metallic particles using the Mie theory because to get a  $\lambda ^{-1}$ law required
smaller particles than if they were dielectric and computations for large particles were too tedious. .

     In 1948 Whitford published measurements of star colors versus spectral types over a
wavelength range from about 3500{\AA} (ultraviolet) to the near infrared. The relation was not the
expected straight line, but showed curvature at the near ultraviolet and infrared regions. Things were
beginning to make some physical sense from the point of view of small particle scattering.

     In the 30's Oort took another approach to the problem by looking at the statistics of the
motions of K giants perpendicular to the plane of the galaxy, that is, at bulge objects. He used these to
estimate the mass of material in the plane. He found that there had to be more material there than could
be seen in stars. Oort and others estimated that the mass of the nonstellar material is about 
$12\times10^9$M$_\odot$. If this mass is distributed uniformly
      $\rho_{ism} \approx 6 \times 10^{-24}  g/cm^3$, 
this is the mass required to explain the observed motions.

     The question then becomes what kind of material distributed with this density with what mass
absorption coefficient could give rise to an optical depth of $\tau/L$ of about 1 magnitude kpc$^{-1}$, as
observed. So what is required is that the scattering/extinction cross section of the material blocking the
light per unit length is on the order of 1 magnitude per kiloparsec. 

     In 1935 Lindblad published an article in Nature that indicated that interstellar abundances made
it seem reasonable to grow particles in space. Eddington long before hypothesized that it was so cold in
space that anything that hits a small particle will stick.

     In the 40's van de Hulst (1949) broke with tradition and published the results of making
particles out of atoms that were known to exist in space: H, O, C, and N. He assumed these atoms
combined on the surface to form frozen saturated  molecules. This is what later became known as the
``dirty ice'' model. The dirty ice model of dust by van de Hulst was a logical followup of the then existing
information about the interstellar medium and contained the major idea of surface chemistry leading to
the ices H$_2$O, CH$_4$, NH$_3$. But it was not until the advent of infrared astronomical techniques made it
possible to observe silicate particles emitting at their characteristic 10$\mu$m wavelength in the
atmospheres of cool stars that we had the cores on which the matter could form. Interestingly, their
presence was predicted on theoretical grounds by Kamijo (1963). As van de Hulst said, he chose to
ignore the nucleation problem and just go ahead (where no one had gone before) with the assumption
that ``something'' would provide the seeds for the mantle to grow on. What a great guess! So by 1945
we had many of the theoretical basics to understand the sources of interstellar dust ``ices'' but it was not
until about 1970 that the silicates were established. However having a realistic dust model, van de Hulst
developed the scattering tools to provide a good idea of dust properties.

     After the extinction curve had been well established and the inferred particle size, two
investigators (Hall 1949; Hiltner 1949) inspired by a prediction of Chandrasekhar on intrinsic stellar
polarization discovered instead the general  interstellar linear polarization. Magnetic fields were
believed to confine cosmic rays and to play a role in the spiral structure of the galaxy. The implication of
the linear polarization was that the extinction was caused by non-spherical particles aligned by magnetic
fields.

     The scattering of light by small particles was understood to be important to understand how
aligned particles could polarize starlight. The calculations for non-spherical particles comparable to the
wavelength was limited to infinite cylinders (which are only finite in two dimensions). This led to attempt
to study arbitrary particles by microwave analog methods  (Greenberg, Pedersen \& Pedersen 1960)
which have led to using such analog approaches to study much more complex particles as are comet
dust (Hage et al. 1991). This method is still proving to be powerful (Gustafson 1999) as the needs still
outstrip the capacities of computers.

\section{The current state of the art}

     The last 40 years have seen a revolution in the study of interstellar dust. This has been a 
threefold process. First of all, the observational access to the ultraviolet and the infrared brought into focus
the fact that there had to be a very wide range of particle sizes and types to account for the blocking of
the starlight. Secondly, the infrared provided a probe of some of the chemical constituents of the dust.
Thirdly laboratory techniques were applied to the properties and evolution of possible grain materials. It
was the advent of infrared techniques that made it possible to demonstrate conclusively that something
like rocks (but very small, of course) constituted a large fraction of the interstellar dust.

     However the first attempt to find the 3.1$\mu$m feature of H$_2$O was unsuccessful (Danielson et al.
1965). This was, at first, a total surprise to those who had accepted the dirty ice model. However, this
gave the incentive to perform the early experiments on the ultraviolet photoprocessing of low
temperature mixtures of volatile molecules simulating the ``original'' dirty ice grains (Greenberg et al.
1972) to understand how and why the predicted H$_2$O was not clearly present. From such experiments
was predicted a new component of interstellar dust in the form of complex organic molecules, as
mantles on the silicates.

     The present studies of interstellar dust take us into many directions, from chemical evolution of
the space between the stars, to comets, to solar systems. I will try to present an overview of our current
knowledge of the dust which brings many related astrophysical problems to the fore. The questions
posed by our present information and the suggested needs for the future are directed towards the
discussion in the last section. The space age has ushered in some of the most dramatic developments
and will ultimately give us the data we need to understand how solar systems are born out of collapsing
clouds.

Some of the outstanding properties and problems currently considered are:
a) Extinction and polarization, b) Chemical constituents (silicates, complex organic, ices, small
carbonaceous particles and polycyclic aromatic hydrocarbons), c) Sources and sinks, d) Relationship to
comets and other primitive solar system bodies, e) Diffuse interstellar bands, f) Scattering by
non-spherical  and aggregate particles, g)Dust absorption and emission, h) Dust/gas chemistry, i) Cosmic
abundances.

\subsection{Extinction and polarization}

     The main characteristics of the ``average'' extinction curve are well established. A slow and then
increasingly rapid rise from the infrared to the visual, an approach to levelling off in the near ultraviolet, a
broad absorption feature at about  $\lambda^{-1} \approx 4.6 \mu m^{-1}$ 
 and, after the drop-off, a ``final'' curving increase to as
far as has been observed $\lambda^{-1} \approx 8 \mu m^{-1}$. 
We note here that the access to the ultraviolet was only made
possible when observations could be made from space, first by rockets and then by satellites (OAO2,
IUE). An example of the  extinction curve is shown in Fig. 1 with data associated with dates of
discovery. I will not discuss here  many important variations which are correlated with different regions
in space. The strength of the hump and the far ultraviolet extinction can vary both independently and
with respect to the visual extinction. The shape of the visual extinction appears to depend on whether it
occurs in high or low density regions. The general shape of the polarization curve is also well
established. It rises from the infrared, has a maximum somewhere in the visual (generally) and then
decreases towards the ultraviolet. The position of the maximum shifts with the ratio of that to selective
extinction R = A$_V$/E(B-V) in the sense that it moves to longer wavelengths as R increases which is the
effect of increasing the particle size (see Whittet 1992).

\begin{figure}
\centering
\includegraphics[height=3in,width=2in,angle=-90]{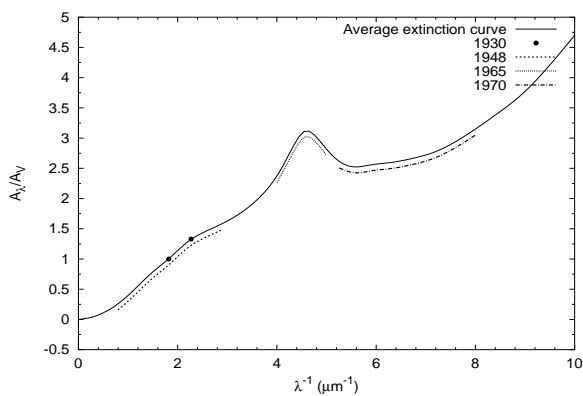}
\caption{How the extinction curve was found. The 2 dots correspond to the color excess derived 
  by Trumpler (1930). The dotted portion corresponds to the Whitford (1948) curve. The hump was 
  discovered using a rocket (Stecher 1965) and the far UV was obtained by the OAO2 and 
  IUE satellites (Bless \& Savage 1970).}
\end{figure}

     The bottom line in the interpretation of the extinction and polarization is that it is  characteristic
of a set of different types of particles spanning a size (``radius'') range from  5nm to  0.2$\mu$m.

\subsection{Chemical composition}

     All of the chemical constituents of interstellar dust derive from infrared observations, with one
exception --- the hump at 220 nm. These constituents are naturally divided into: 1) volatiles, 2)
refractories. The established ``solid'' chemical constituents of interstellar dust are silicates, carbonaceous
material and various frozen volatiles, commonly called ices. Another component whose presence
appears highly likely exists in the form of large polycyclic aromatic hydrocarbons (PAH's). The
variations in the properties and relative proportion of the different constituents is one of the most
important characteristics of the dust.

\subsubsection{Volatiles}

 The infrared observations of absorption have demonstrated the presence of icy mantles on the
interstellar dust. In Table 1 is presented the kinds and relative amounts of molecules detected in
different regions in space. Not only are the relative proportions variable but also the presence and
absence of some species. In almost all cases, however, water is the dominant component. Another
important variability is in the layering of the various molecular components. Of particular note is the fact
that the CO molecular spectrum is seen to indicate that it occurs sometimes embedded in the H$_2$O 
(a polar matrix) and sometimes not. This tells a story about how the mantles form. As was first noted by
van de Hulst, the presence of surface reactions leads to the reduced species H$_2$O, CH$_4$, NH$_3$. Since we
now know that CO is an abundant species as a gas phase molecule, we expect to find it accreted along
with these reduced species --- at least initially.

\begin{table}
\caption{Ice mantles on dust in clouds and around protostars}
\begin{tabular}{lcc} \hline
  Species	& Protostars	& Field Stars \\ \hline
  H$_2$O	& 100		& 100 \\
  CO		& 1--15 (polar) & 7 (polar) \\
		& 1--50 (apolar)& 27 (apolar) \\
  CO$_2$	& 15--40	& 15 \\
  CH$_4$	& 1--4		& -- \\
  CH$_3$OH	& 1--35		& $<3.4$ \\
  H$_2$CO	& 3		& -- \\
  OCS		& 0.05--0.18	& --	\\
  NH$_3$	& 3--10		& $<6$	\\
  C$_2$H$_6$	& $<0.4$	& --	\\
  HCOOH		& 3		& --	\\
  O$_2$		& $<20$		& --	\\
  N$_2$		& ??		& ??	\\
  OCN$^-$	& 0.3--2.9	& $<0.4$	\\
  HCN		& $<3$		& --	\\
  \hline
\end{tabular}
\end{table}	

     As one can see from comparing the relative abundances of the ice mantle molecules and the gas
phase molecules, the ice mantles are not made simply by accretion from the gas. An
obvious case in point is the H$_2$O:CO ratio which is so much higher in the dust than in the gas. Also
notable is the fact that in star forming regions the relative abundance of H$_2$O and CH$_3$OH is very
different.

     The two approaches to understanding how the grain mantles evolve are: (1) the laboratory
studies of icy mixtures, their modification by ultraviolet photoprocessing and by heating, (2) theoretical
studies combining gas phase chemistry with dust accretion and dust chemistry. In the laboratory one
creates a cold surface (10K) on which various simple molecules are slowly deposited in various
proportions. The processing of these mixtures by ultraviolet photons and by temperature variation is
studied by infrared spectroscopy. This analog of interstellar dust mantles is used to provide a data base
for comparison with the observations. Figure 2 shows a typical observation with appropriate molecular
identification based on the laboratory.

\begin{figure}
\centering
\includegraphics[height=3in,width=2in,angle=90]{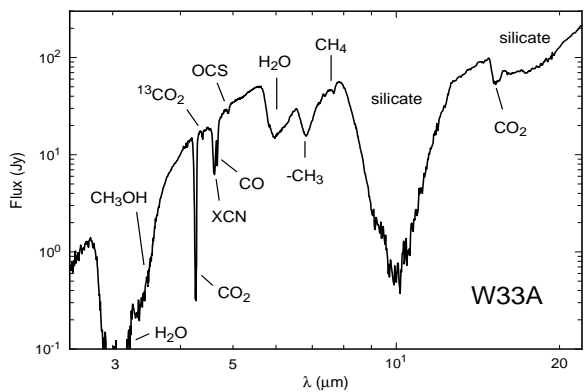}
\caption{Spectrum of the dust around the embedded protostar W33A showing identified molecules 
 in the ice mantles (Gibb et al. 1999). XCN is identified as OCN$^-$ (Schutte \& Greenberg 1997).}
\end{figure}

\subsubsection{Refractories --- sources and sinks}

The most ubiquitous refractory dust component is indicated by the interstellar 10 $\mu$m absorption feature.
It is ascribed to the Si-O stretch in some form of rock. The shape of the interstellar feature is broad and
featureless both of which suggest that the silicate is amorphous. The originating source of the interstellar
silicates is in the atmospheres of cool evolved stars where the emission features clearly indicate a
fractional degree of crystallinity (Waters et al. 1996). How can interstellar crystalline silicate particles become
amorphous? This is a puzzle which has not yet been completely solved.

     The second refractory dust component is generally referred to as carbonaceous. It appears to
occur in three generic forms: (1) as mantles on the silicates, (2) as small particles which produce the
220 nm hump, (3) as large polycyclic aromatic hydrocarbons (these are often considered also as
particles). 

     What is currently known about all of these components is based very largely on results of
laboratory experiments which attempt to simulate interstellar processes. The organic refractory dust
mantles which are derived from the photoprocessing of ices contain a mixture of aliphatic and aromatic
carbonaceous molecules. The laboratory analog suggests the presence of abundant prebiotic organic
molecules in interstellar dust (Briggs et al. 1992). An excellent match to the observed infrared 3.4 $\mu$m
feature (Greenberg et al. 1995) is provided by photoprocessed laboratory organics and there is
indication of absorption in the 8 - 10 $\mu$m region which satisfactorily modifies the silicate absorption
feature to match observations (Greenberg \& Li 1996). However, hydrogenated amorphous carbon
made by condensing carbon smoke also matches the 3.4 $\mu$m feature. We need
more definitive laboratory experiments and astrophysical theories to derive the chemical composition of
the organic mantles as well as the other carbon based materials.

     The hump particles seem to be best described at present as resulting from dehydrogenation of
hydrogenated amorphous carbon and the PAH's may well result from dehydrogenation of residues from
photoprocessed ices (Mennella et al. 1998). What is being indicated here is the fact that, just as ices
evolve chemically and physically in interstellar space, so do the organics. 
Where and how the interstellar dust is formed appears to involve a complex evolutionary picture. The
rates of production of refractory components such as silicates in stars do not seem to be able to provide
more than about 10\% of what is observed in space because they are competing with destruction which
is 10 times faster by, generally, supernova shocks. At present the only way to account for the
observed extinction amount is to resupply the dust by processes which occur in the interstellar medium
itself. The organic mantles on the silicate particles must be created at a rate sufficient to balance their
destruction. Furthermore, they provide a shield against destruction of the silicates. Without them the
silicates would indeed be underabundant.

\subsection{Interstellar dust and comets}

\begin{table}
 \caption{Comparison of the abundances of ices in the interstellar medium 
          (towards IRS9) and of cometary volatiles (at $\sim$ 1 AU). Cometary
	  abundances are from Table 7 in Crovisier (1999). ISM ice abundances
	  are taken from a compilation of Schutte (1999) from various sources
	  as well as from recent \emph{ISO} results (d'Hendecourt et al. 1996,
	  Whittet et al. 1996)}
 \begin{tabular}{lclcl} \hline
  Species &\multicolumn{2}{l}{Interstellar ices} &
    \multicolumn{2}{l}{Cometary volatiles}	\\ \hline
  H$_2$O	& =100	 &		& =100		&	\\
  CO		& 10--40 &		& 2--20		&	\\
  CH$_3$OH	& 5	&		& 1--7		&	\\
  CO$_2$	& 10	&		& 2--6		&	\\
  H$_2$CO	& 2--6 & tentative	& 0.05--4	&	\\
  HCOOH		& 3    & tentative	& $\sim$ 0.1	&	\\
  CH$_4$	& 1--2		&	& 0.7	&		\\
  other hydrocarbons & ??	&	& $\sim$ 1 & C$_2$H$_2$, C$_2$H$_6$ \\
  NH$_3$	& $<10$		&	& 0.5			\\
  O$_3$		& $\leq$ 2	&	& ??			\\
  OCN$^-$	& $\leq$ 0.5--2	&	& 0.2 & nitriles + HNCO	\\
  OCS		& 0.2		&	& 0.4 & OCS + CS		\\
  SO$_2$	& ??		&	& $\sim$ 0.1	&	\\
  H$_2$		& $\gtrsim$ 1	&	& ??	&		\\
  N$_2$		& ??		&	& ??	&		\\
  O$_2$		& ??		&	& ??	&		\\ \hline
 \end{tabular}
\end{table}

     A major advance in our understanding of comets in the 20th century was made by the space
probes Vega 1 and 2 and Giotto (see Nature, comet Halley issue, vol 321, 1986). 
Until that time no one had ever seen a comet
nucleus. The critical new discoveries were: (1) the low albedo of comets, (2) the size distribution of the
comet dust extending down to interstellar dust sizes (10$^{-15}$ -- 10$^{-18}$ g), (3) the organic fraction of comet
dust. The current ground based observations of the volatile composition of comets implies a close
connection with the ices of interstellar dust. This is shown in Table 2. Most of the current models of
comet nuclei presume that to a major extent they are basically aggregates of the interstellar dust in its
final evolved state in the collapsing molecular cloud which becomes the protosolar nebula. In addition to
the chemical consequences of such a model there is the prediction of a morphological structure in which
the aggregate material consists of tenth micron basic units each of which contains (on average) a silicate
core, a layer of complex organic material, and an outer layer of ices in which are embedded all the very
small carbonaceous particles characterizing the interstellar ultraviolet hump and the far ultraviolet
extinction  All these components have been observed in the comet comae in one way or another. The
implication is that space probes which can examine in detail the composition of comet nuclei will be able
to provide us with hands-on data on most of the components of interstellar dust and will tell us what is
the end product of chemical evolution in a collapsing protosolar molecular cloud. At this time many
laboratories are preparing materials as a data base for comparison with what will be analyzed during the
space missions.

\subsection{Dust absorption and emission}

     One of the achievements of Van de Hulst (1949) was the calculation of the temperature of
interstellar dust size particles. Because of their small size they absorb well at short wavelengths but
poorly at long wavelengths. Because of the fact that their general environment is in a radiation field
which, while having a high color temperature (high energy ultraviolet photons) is of low energy density,
they are cold but not as cold as a black body; i.e., characteristically 
T$_d \approx$ 15K rather than T$_{BB}$  = 3.2
K. While later refinements in this by Greenberg (1971) confirmed this, the advent of the space infrared
measurements by the IRAS seemed to detect only about 10 -- 20 \% of the dust. This was simply a
result of a lack of a sufficiently long wave length detector to see the cold dust. It took an entirely
indirect method developed by Block (see review by Block 1996), where the spiral arms of galaxies
provided the tracers of low temperature dust accounting for 80 -- 90 \% of the total mass and, in fact,
this is the dust which produces all the visual extinction and the infrared absorption features. The
complete emission spectrum by the dust of all sizes provided a good comparison with satellite
observations (COBE, DIRBE) in the far IR and submm --- as well as the near to mid IR (Li \& Greenberg
1998a,b). Thus space observations have proven to be an absolute requirement.

\subsection{Dust/gas chemistry}

     There is inadequate space here to fully document the immense progress which is being made in
this field. For a recent review see Van Dishoeck (1999). I believe that this will be a field which will see
even deeper application in the next century. The principal theoretical problems remaining have to do
with atom and ion surface reactions and desorption mechanisms from the dust. These problems will
only be solved by new experimental  and observational techniques.

\subsection{Cosmic abundances}

     Derivations of the relative abundances of the elements in our galaxy are one of the principal
needs for understanding the chemical evolution in interstellar space - and ultimately its memory in
comets. A major factor in developing consistent dust models was the observation of the ``depletion'' in
low density clouds using the ultraviolet as a probe. The deduced possible dust composition was initially
only constrained to the extent that silicates alone could not be responsible for the interstellar extinction
(Greenberg 1974). But In Recent Years, The problem of grain modeling has been exacerbated by the
apparent decrease of the available condensible atoms (O, C, N, Si, Mg, Fe) by about 30 \% since the
solar system was born (see Mathis 1996; Li \& Greenberg 1997). This implies that the heavy elements
are being consumed more than they are being created. However, if one goes back far enough in time,
there were no condensible atoms because their initial production must follow the birth of stars. This
brings us to the cosmological question of what do high z galaxies look like and when and how was dust
first found in them?

\section{Future}

     There are quite a few unsolved or partially solved problems related to interstellar dust which
will be demanding close attention in the future:

\subsection{Some outstanding problems}

     1)   How does dust evolve in protosolar regions? We need higher spatial resolution and sensitivity.
     Improvements in the theory of dust/grain chemistry, particularly in collapsing clouds leading to
     star formation as well as in quiescent molecular clouds. How do interstellar grains accrete and
     deplete mantles in dense molecular clouds? We need high spatial resolution observations of
     molecule distributions in the gas and in the solid as function of depth in the cloud - interiors of
     clouds as well as regions of low and high mass star formation.

     2)   What is the true atomic composition of the interstellar medium? How variable is it in time and
     space? Are there global variation over distances of kiloparsecs? When did dust first form in a
     galaxy?

     3)   What is the source and nature of the DIBs?

     4)   Will the chemical and morphological analysis of comet nuclei and dust material reveal the true
     character of interstellar dust? Will they provide further answers to the question of life's origin?

     5)   How can we resolve the evolution of interstellar matter leading to the material measured and
     analyzed in meteorites? in interplanetary dust particles?

     6)   What are all the sources and sinks (destruction) of interstellar dust?

\subsection{Upcoming observations}

     To give some impression of some of the exciting possibilities the following is a list of new
remote observational facilities which will be available early in the new millennium.
They are largely in the
longer wavelength regions of the spectra.

     a)   Stratospheric Observations Far Infrared Astronomy (SOFIA) 
          2.5m telescope in a B747 with state of the art instruments and resolution up to R $\approx$ 20,000

     b)   Space Infrared Telescope Facility (SIRTF) 
            60cm cooled telescope R $\approx$ 600, 10--38$\mu$m

     c)   Next Generation Space Telescope 
          8m passively cooled telescope  0.6 -- 10 $\mu$m (up to 30 $\mu$m);  R $\lesssim$ 3000

     d)   Submillimetre Wave Astronomy Satellite (SWAS)
           90cm;  557/484  GHz : H$_2$O, O$_2$, CI; R$>$10$^6$ (heterodyne spectroscopy)

     e)   Far Infrared and Submillimeter Telescope (FIRST)
          3.5m;  80 -- 500 $\mu$m;  R $\gtrsim$ 10$^6$  (heterodyne spectroscopy)

     f)   Atacama Large Millimeter Array (MMA/LSA)
          64$\times$12m telescopes  0.3 -- 3$\mu$m; 80 -- 800GHz  R $\gtrsim$ 10$^6$ (heterodyne spectroscopy)

\section{Cosmic Dust and Origins}

     There are many implications of future cosmic dust studies by new and improved methods.
Three, which are very fundamental and most exciting have to do with origins:

(1) the early universe

(2) the origin of the solar system

(3) the origin of life

\subsection{High-z galaxies}
 
    How and when did dust first appear in our young galaxy? How rapidly did the condensible
elements form and what was the nature of the first kind of dust? The earliest heavy elements could not
have been created until after the first stars and the first supernovae. The only access we have to answer
these questions is from the infrared emission and obscuration by dust in high-z galaxies.   Such galaxies
with far infrared (FIR) luminosity $>10^{12}$L$_\odot$   were among the most important discoveries made with
IRAS. These must represent galaxies undergoing massive rates of star formation. In order to observe
the rest frame emission by cool dust (accounting for the major dust mass) in such high-z (2-5) galaxies
we have to obtain multiwavelength data in the submillimeter to millimeter region so as to span the range
of possible spectral emission turnovers between 350$\mu$m and 850$\mu$m. In our own galaxy this would be
about at 150$\mu$m but not only is the newly formed dust unquestionably not like ours but the absorbed
stellar emission is at much higher temperatures and densities in these star burst galaxies.
How rapidly does the dust evolve in very young galaxies? The fact that there are even some differences
between the dust in the Magellanic clouds and in our Milky Way (hereafter MW) 
gives us cause to seriously consider
major differences between the present MW  dust 
and the first dust. In the Magellanic clouds the extinction
rises monotonically in the UV with no hump. It will be a real challenge to
model early dust and to do so, we will need the rest infrared emission in great detail because it is the
only property we can observe---no extinction curve, no chemistry, etc. The earliest dust must have had a
hard time because both the source of the heavy elements to make the dust and the destruction of the
dust are by supernova. The maximum dust size was severely limited. Would tenth micron particles exist as in
our galaxy? How early did late type (low mass stars) evolve to produce silicates?  Although M
supergiants occur early on in a very young galaxy their rates of silicate production---if comparable with
that of current M supergiants---would be counterbalanced by a much greater supernova destruction
rate. For example, it appears that the supernova production rate in a z = 3.8 galaxy is at least 3 more orders
of magnitude higher than in our galaxy (Dey et al. 1997). It would be hard to see how the silicate dust
at that epoch could have been larger than, say a few hundredths of a micron. The temperature and
extinction properties would be totally unlike that of interstellar dust as we know it. The ultraviolet
extinction would be an order of magnitude larger than visual extinction rather than a factor of 2 as in MW
dust (see Fig. 1 and Fig. 7). In order to begin to see early galactic dust bearing any resemblance to ours I would
expect one would have to wait until two things occur: the supernova rate drops precipitately and the M giants
(low mass stars) become abundant---perhaps several billion years. We would need the time to grow
mantles on the dust if larger particles are to survive.

We have thought it useful to present a preliminary set of results for extinction and emission based on the
idea that the dust which first appears in high-z galaxies is indeed not like what is observed in later
epochs. We start with the assumption that the major first source of dust is from supergiants for which
the timescale for appearance may be only $10^6$---$10^7$ yr. In our galaxy the major source of silicates
appears to be predominantly M giants but these low mass stars will have taken longer to evolve and it is
only at the time they appear that the dust may begin to resemble more what we see in our galaxy with
the accretion of mantles. At z as high as 2.5 and 4.69, CO has indeed been 
detected (Scoville et al. 1996; Omont et al. 1996)
and if this is an indication of the possible existence of grain mantles then one may perhaps observe dust
more like ours but we note that CO has not been detected in a z = 3.4 radio galaxy B2 0902+34 
(Eales et al. 1993). However, we should point out that even with CO present, the accretion of atoms
and molecules would be strongly inhibited by the high grain temperatures which, as we shall see, are
considerably higher than the value of T=15K characteristic of the MW because of the massive star
formation rates. Mantle growth time in our galaxy assuming standard cosmic abundances of 
(O+C+N)/H $\approx 10^{-3}$ and a cloud density of n$_H \approx 10^4$ cm$^{-3}$ 
is $\sim 5 \times 10^7$ yr. The lifetime of dust fully exposed to
supernova shocks is $< 5 \times 10^7$ yr even for S.N. formation rates 
only 10---100 times more than the 0.03 yr$^{-1}$
in our galaxy .  Dey et al. (1997) suggest as many as $10^3$ supernova yr$^{-1}$ as not unrealistic in high-z
galaxies. Not until the supernova rates drop and M giants (from relatively low mass stars) begin to appear can
we expect the dust to begin to resemble what we call ``normal'' galactic dust. Thus galaxies with
luminosities much higher than the MW may well be observed in CO but may not have 
MW type dust. With this proviso we now assume that the dust in the (sufficiently) high-z galaxies is
limited to the production  by supergiants; namely, silicates. The sources of carbon particles are
presumed to be less productive and we also presume that supernova dust production  is --- as in our
galaxy --- less than that of the supergiants.

     We shall calculate the dust spectral energy distribution (SED) for a variety of parameters which
span possible conditions in early galactic evolution. With very high large mass star formation rates we
consider the possibility that the mean radiation field to be characterized by a temperature as high as T$_R$
= 30,000 K as compared with the T$_R^{MW} \approx$ 10,000 K (mean A star source). Depending on the stellar
density we consider dilution factors ranging from W = 10$^{-14}$ (as in the MW) to as high as W = 10$^{-12}$
representing an extreme case. The combination W = 10$^{-12}$ with T$_R$ = 30,000 K provides a radiation
energy density $\sim 10^4$ that of the M.W.

\begin{figure}
\centering
\includegraphics[height=3in,width=2in,angle=-90]{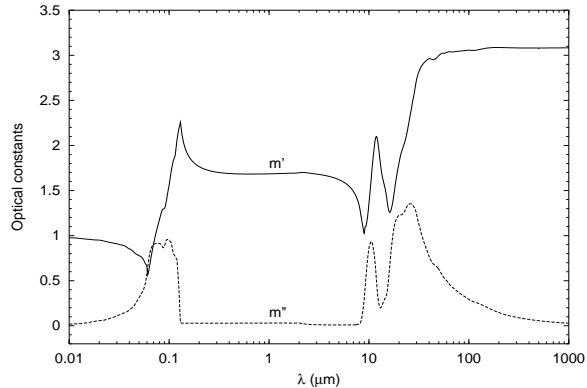}
\caption{Optical constants m($\lambda$)=m$'$($\lambda$)-im$''$($\lambda$) of silicates from Li \& Greenberg (1997).}
\end{figure}

     The particles are taken to be pure silicates with optical constants as shown in Fig. 3. They are
presumed to have a mean size (radius) \={a} = 0.05 $\mu$m which we believe to be characteristic of the
particles made in the evolved star envelopes seen in our galaxy (Tinbergen et al. 1981). To take into
account the dependence of the emission on grain size we consider a size distribution:  $n(a)\sim a^{-3},
0.03 < a < 0.1\mu m$, with the upper limit estimated to be the maximum size possible in the expanding stellar
envelope. The average of this distribution is 0.05 $\mu$m. The grain temperatures are calculated by equating
the absorbed and emitted radiation
\begin{equation}
   \int \pi a^2 Q_{abs}(a,\lambda)R_G(\lambda)d\lambda = \int \pi a^2 Q_{abs}4\pi B(\lambda,T_d)d\lambda
\end{equation}

\begin{figure}
\centering
\includegraphics[height=3in,width=3in]{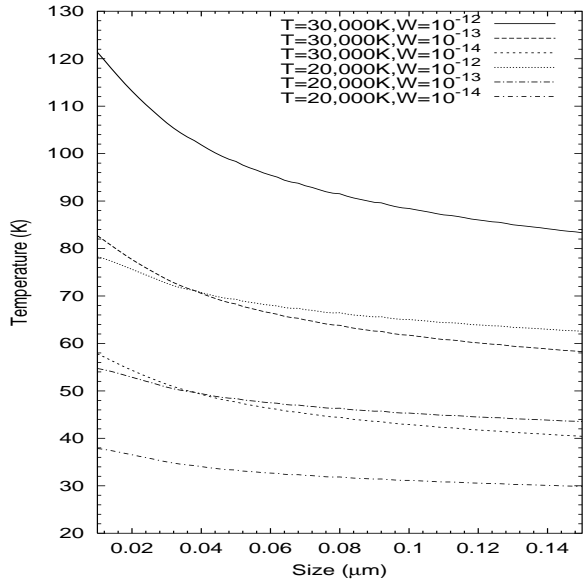}
\caption{Grain temperature as a function of size for various radiation fields as specified
 by a temperature T and a dilution factor W.}
\end{figure}

where Q$_{abs}$ is the Mie theory absorption efficiency, R$_G$ is 
the assumed galactic radiation field, B($\lambda,T_d$) is
the black body radiation at the dust temperature T$_d$. Fig. 4 shows the dust temperature as a function of
size for different radiation fields.  These temperatures are generally in the range expected (see e.g, 
Isaak et al. 1994). We see that for these pure silicate particles the temperature monotonically decreases
with increasing size --- a result which differs from that for similar particles in the solar radiation field
(Greenberg \& Hage 1990). As a convenient tool we show the emission and absorption by a 0.05 $\mu$m
silicate particle in Fig. 5 in which it is demonstrated how to deduce readily the temperature for different
radiation fields and dilution factors. 

\begin{figure}
\centering
\includegraphics[height=3in,width=2in,angle=-90]{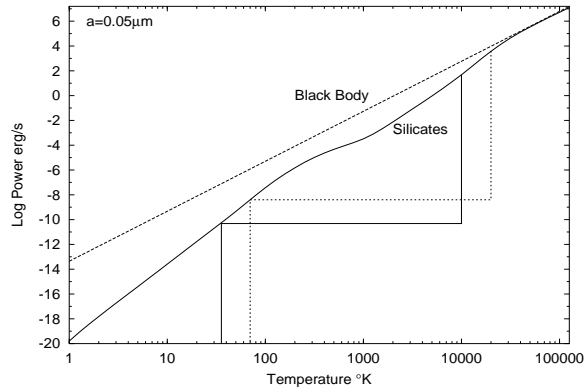}
\caption{Emission/absorption of a 0.05$\mu$m size silicate particle as
 a function of temperature compared with a black body. As examples of how
 to use this to obtain a grain temperature see the demonstrations for T=10,000K and T=20,000K,
 W=10$^{-12}$ by the sequences of vertical, horizontal and vertical lines.}
\end{figure}

\begin{figure}
  \centering
  \includegraphics[height=4in,width=2in,angle=-90]{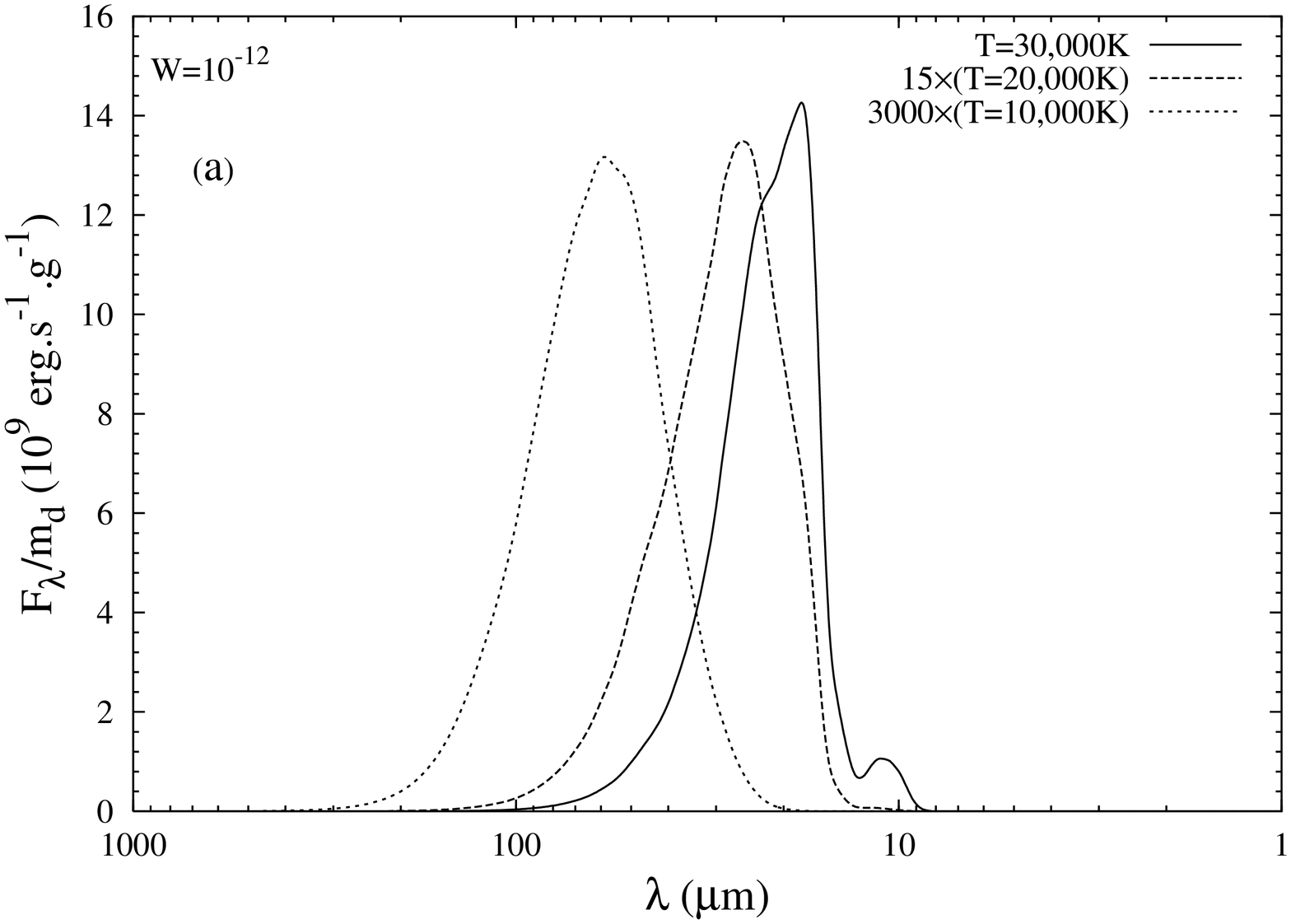}
  \includegraphics[height=4in,width=2in,angle=-90]{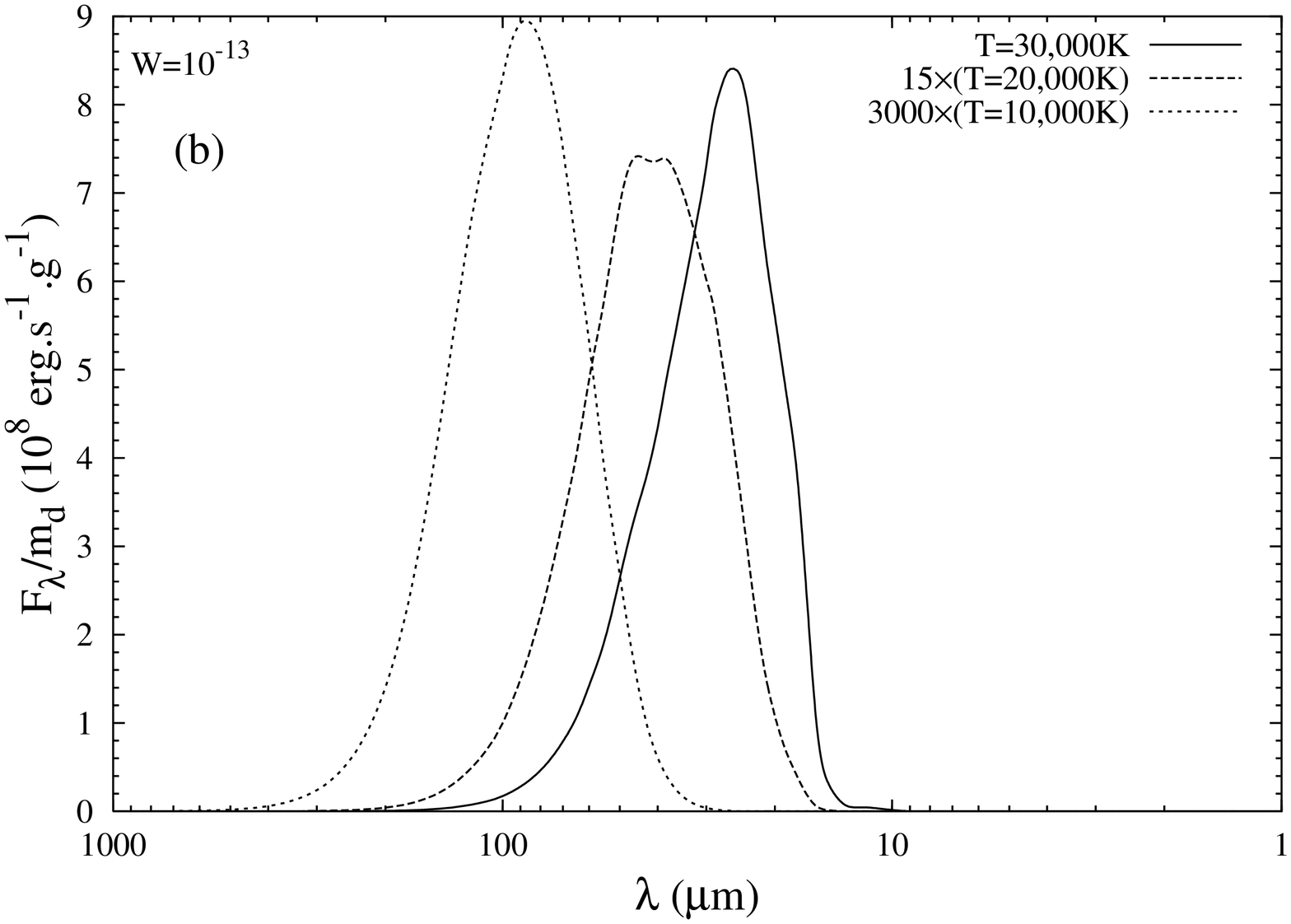}
  \includegraphics[height=4in,width=2in,angle=-90]{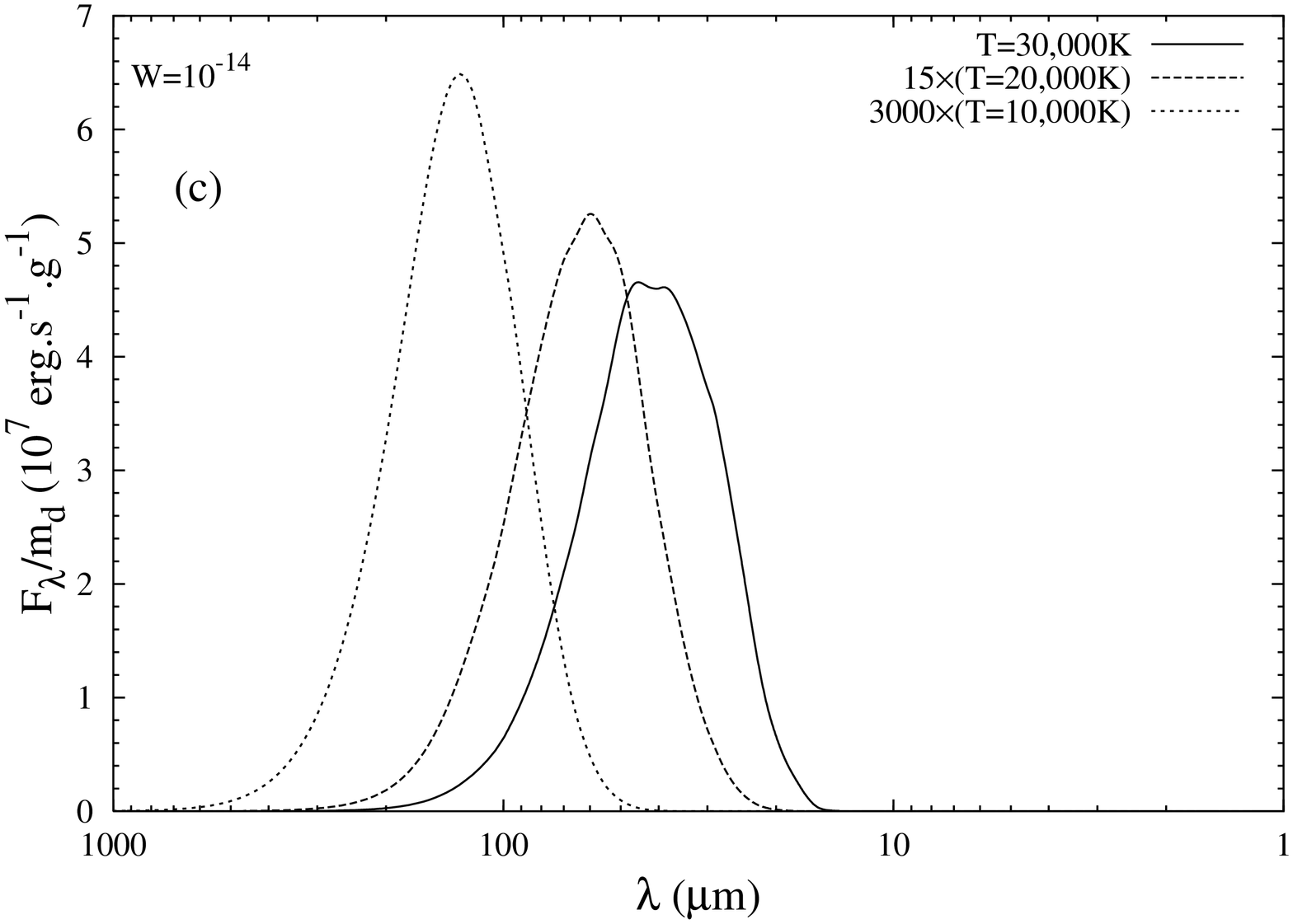}
  \caption{Thermal emission per unit mass for silicate dust grains  with 
  mass density 3 g.cm$^{-3}$ and mean size \={a}=0.05$\mu$m in radiation field (T)
  and dilution factor (W). The size distribution of grains is $n(a)\sim a^{-3}$, 
  $0.033<a<0.1\mu$m, and: a) W=10$^{-12}$, b) W=10$^{-13}$, c) W=10$^{-14}$. Note 
  that T=10,000K, W=10$^{-14}$ correspond approximately to the Milky Way 
  radiation field.}
\end{figure} 

     In Figure 6a,b,c we show the emission per unit mass for a variety of radiation fields with an
assumed mean silicate density of 3 g$\cdot$cm$^{-3}$. Note that we have used the approximation 
that the dust is uniformly distributed in the mean
radiation fields of the galaxy. For such a case the extinction will be variable in the  galaxy --- stars in the
center being more obscured than those in the outer regions. It is well recognized that the total IR flux
from the homogeneously mixed stars and dust is expected to be greater than that in which the extinction
is less (see Calzetti 1999a,b and references therein).

 \begin{figure}
  \centering
  \includegraphics[height=3in,width=2in,angle=-90]{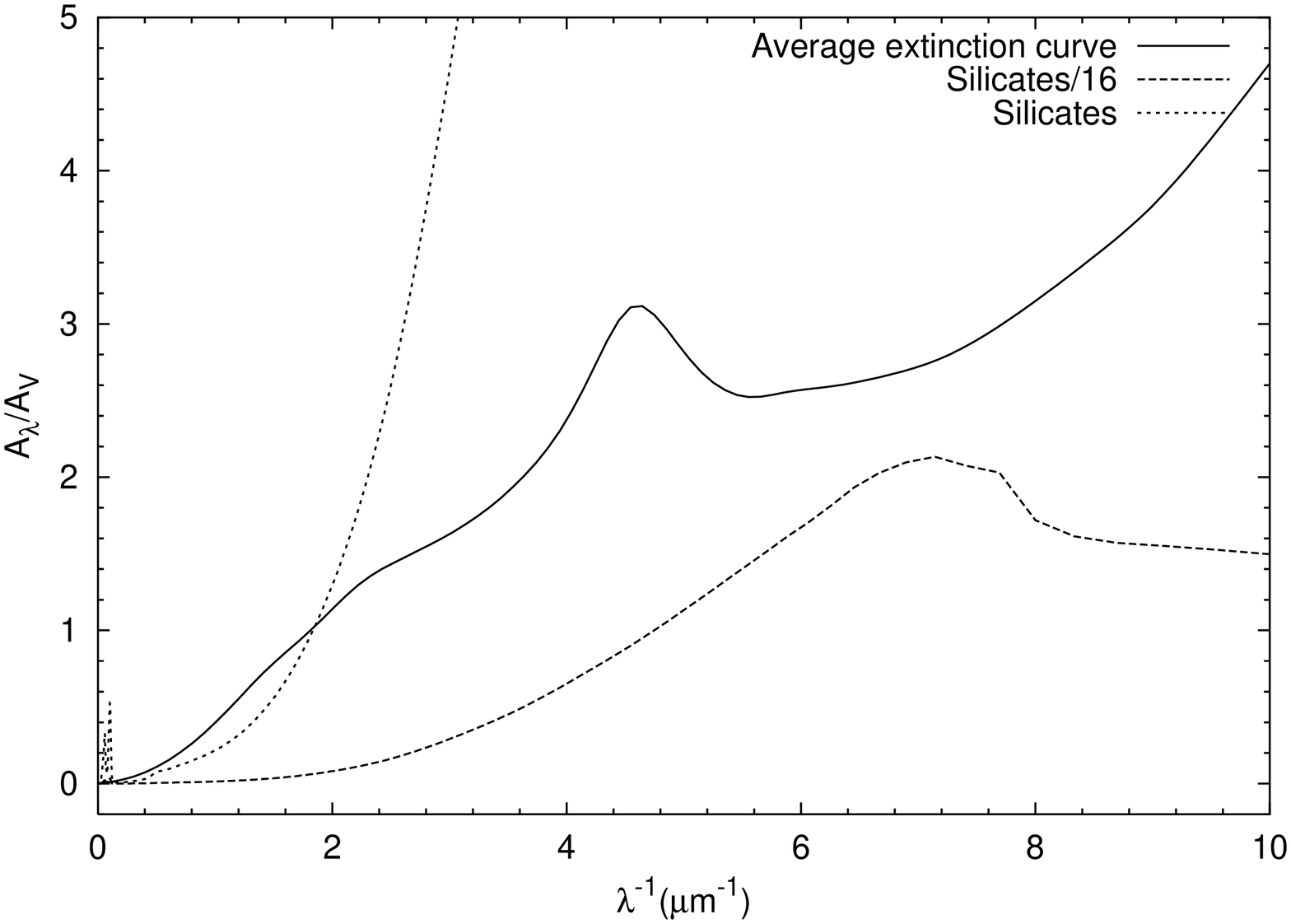}
  \caption{Extinction curve for a size distribution of silicate particles
    ($n(a) \sim a^{-3}, 0.033 \leq a \leq 0.1\mu m$) with mean size \={a} = 
    0.05$\mu$m. Upper dotted curve nomalized to A$_V$=1, and lower dashed curve
    divided by 16. Extinction per unit mass at V is 
    about $9.4\times 10^3$ cm$^2\cdot$g$^{-1}$.}
 \end{figure} 

     A major effect introduced by the dust model considered here is in the extinction curve. In Fig. 7
we present the extinction as if the dust is distributed in a shell (the screen model). This
overestimates the extinction but it is the wavelength dependence which we think is of great interest.
Even if multiple scattering effects were taken into account, the ratio of UV to visual extinction is about
ten times larger than that for the ``standard'' MW average diffuse cloud extinction curve. We have
not yet explored whether this degree of anomaly in the extinction curve is consistent with observations
but we believe that in any case the energetics of the high-z galaxies should prevent the mean dust size
being as large as in the evolved galaxies so that the UV to visual extinction is expected to be large.
We note that the extinction per unit mass at V for the \={a}=0.05$\mu$m silicates is only 1/5 that of 
silicate paticles large enough ($a\approx 0.1\mu$m) to give a more ``normal'' extinction curve. The extinction
per unit mass is presented over the full wavelength in Fig. 8.

 \begin{figure}
  \centering
  \includegraphics[height=3in,width=2in,angle=-90]{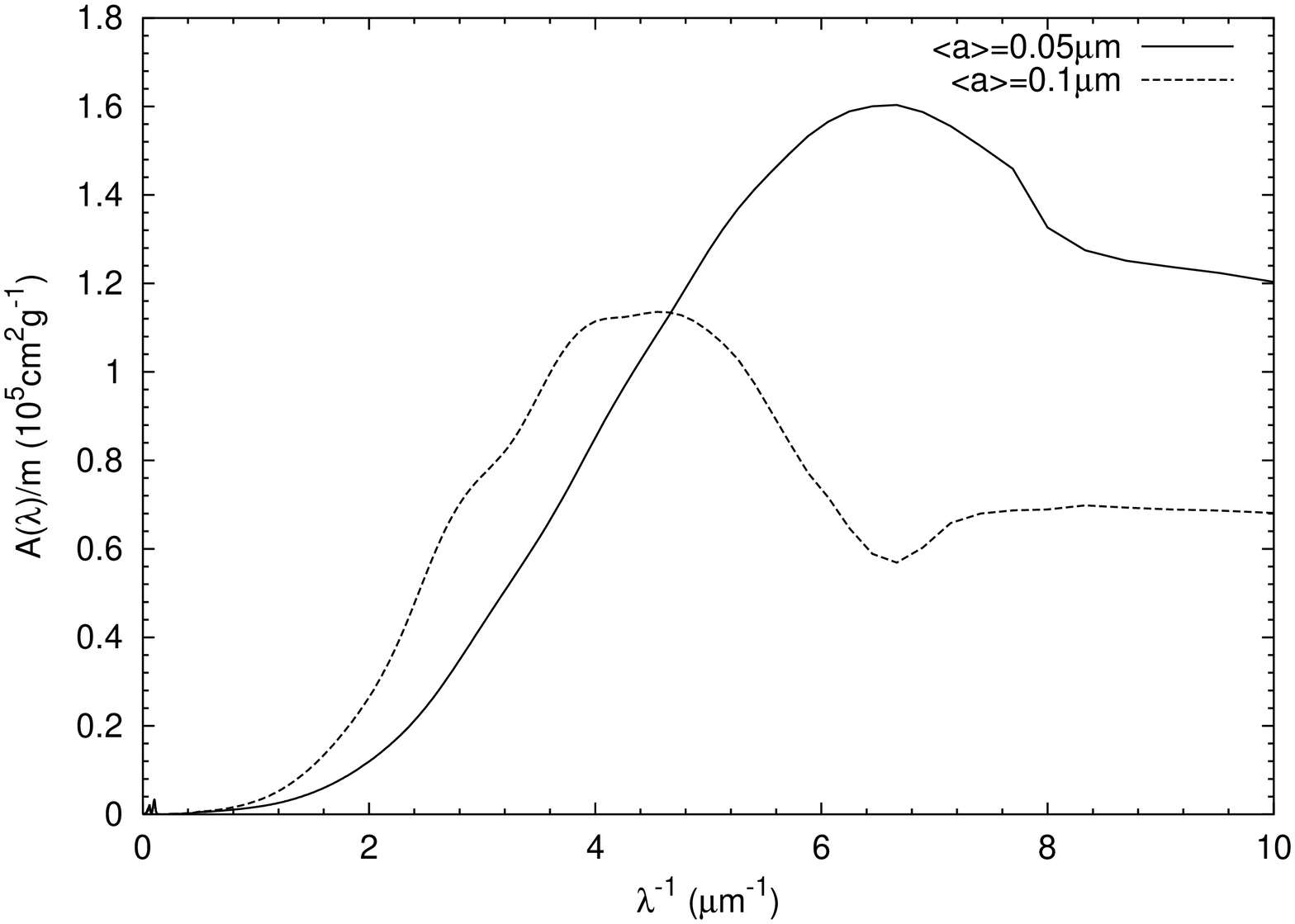}
  \caption{Extinction per unit mass, A($\lambda$)/m (=1.086$\times \tau$/m), for a size distribution
    of silicate particles, $n(a)\sim a^{-3}, 0.033 \leq a \leq 0.1 \mu$m with a mean size
    \={a}=0.05$\mu$m and for $0.09\leq a \leq 0.113\mu$m with a mean size \={a}=0.1$\mu$m.}
 \end{figure} 

     Does the anomaly in the extinction curve we calculate help to answer the question raised by
Calzetti (1999b) ``What is causing the UV stellar continuum of Lyman-break galaxies to be redder than
expected for a young star-forming population?''  It is certainly true that the dust correction in the UV
implied by the curve in Fig. 7 is greater for the same dust mass than would have been expected.

The long term future of high z astronomy at submillimeter wave length lies with interferometric
observation with further gain in sensitivity and angular resolution which will make it possible study single
protogalactic structures at high z. The next generation of bolometer arrays such as SCUBA on the
JCMT will go far in this direction.

A theoretical study of emission properties of postulated early dust must consider possibilities very
different from what we ``see'' in our Milky Way and neighboring galaxies. The evolution of earliest dust is
likely to be very different from that in our relatively quiescent Milky Way so that the larger particles may
not ever accrete mantles of, say, ices or carbonaceous material. Suppose they are purely silicates or
carbon or SiC? Their temperatures and the dust/gas ratio may be quite difficult to reconcile with our
preconceived notions.

\subsection{Origin of the Solar system}

     The most pristine relics of the early solar system are comets. From the first space missions to
comet Halley we already became aware of their complex chemistry and, as summarized in section 3.3,
their undoubted resemblance to an aggregate of interstellar dust. Both the US NASA and the the
European ESA have planned missions to comets and asteroids which will reveal even more details
about their composition. It is the goal of the ESA Rosetta mission to explore the nucleus of a comet
(P/Wirtanen) in detail from close distance over an extended period of time not only remotely from the
Orbiter, but also by means of a Lander on the surface of the nucleus (Verdant \& Schwehm 1998).
What were the conditions in the solar system when comets were formed? How was the collapsing
cloud material affected by heating and turbulence? One of the aspects of comets that will tell us about
these conditions is the density. Remarkably, as much as we have learned about comets there are still
debates about their density. The Rosetta orbiter will both see the comet and it will be able to measure
its mass from its orbit. Thus, for the first time we will know whether comets are as fluffy as has been
predicted. (Greenberg \& Hage 1990; Li \& Greenberg 1999). This information is critical to appreciate the
initial aggregation processes which occurred in the pre-solar nebula. The orbiter will follow the comet
from the time it is about 4.5AU until about 1AU. During this time, the COSIMA (Kissel, 1999)
instrument will capture the dust and analyze it for both its organic and inorganic constituents, including
isotopic abundances. Far from the sun the cometary particles are expected to contain many of the ices
remaining from the original interstellar dust. Farther in, the remaining constituents will be the complex
organic and the rocky/metallic components.

The molecular composition of the organic phase of the solid cometary particles will be used to establish
its exobiological relevance as possible organic precursors to the origin of life. (See next section).
The COSAC experiment on the lander (Rosenbauer et al. 1999) is also aimed at the in situ investigation
of the matter of a cometary nucleus with respect to its chemical and isotopic composition.
In addition it will measure its chirality (handedness) of the organics to answer the question of whether
the left handedness of the amino acids of life on earth originated from the interstellar dust which
aggregated to form comets.

Not only will the chemistry and density be measured but also the structure. The grain morphology will
be studied with an atomic force microscope with nm resolution. This should tell us something about the
size of the individual postulated interstellar dust particles.
Considering what will be achieved in the next 20 years I would venture the prediction that within
another couple of decades we will have the capability to ``mine'' pristine comet nucleus material and
bring it back to the earth for laboratory analysis. If comets are indeed made of interstellar dust, we will
have in our hands the ultimate form of molecular cloud dust resembling the dust around low mass
protostars-4.6 billion years ago. We will know the material out of which the solar system was formed.

\subsection{Origin of Life}
 
     It has become generally accepted that the complex molecules required to initiate life on earth
very likely came from outer space rather than having been formed on the earth. In any case, the
abundance of complex organics in interstellar dust as inferred from infrared spectra and from laboratory
analog experiments is certainly an immense reservoir of prebiotic molecules. There is reason to
believe that the interstellar dust in the presolar nebula had an excess of 
lefthanded amino acids (Cronin \& Pizzarello 1997; Engel \& Macko 1997). It is
these molecules---as a major fraction of the interstellar dust--- which makes up about 20\% of the mass of a
comet (Greenberg 1998). If comets are indeed very fluffy aggregates of 
the interstellar dust (Greenberg 1986), the early
bombardment of the earth by comets brought an enormous amount of complex organics to the earth in
the form of very low density comet dust particles. Favorable factors provided by comet dust to the
initiation of life involve three factors: (1) high porosity, (2) inorganic (mineral) surfaces for catalysis, and
of course, (3) the presence of preformed chiral organics. We envisage that these comet dust fragments
may fulfill the requirements suggested by Krueger and Kissel (1989) for chemical thermodynamics to
start molecular self-organization. Each comet dust particle consisting of a porous aggregate of submicron
silicate cores with organic refractory mantles sitting within a water bath (the local ``ocean'') containing
nutrients satisfies the conditions for non-equilibrium thermodynamics as in Nicolis and Prigogine (1977).

     The discussion by Krueger and Kissel (1989) points out particularly to the chemical composition
of the organics predicted by Greenberg (e.g. Briggs et al. 1992)---nitrogen as a hetereroatom and as
reactive precursors of purines and pyrimidines-possible cytosine. The oxygen is present in aldehydes and
carboxylic acids, carbon oxides and of course, water. Most of the hydrocarbons are unsaturated and the
aldehydes may polymerize to form sugars given the right conditions.

What might we anticipate being accomplished in the next millennium? When we retrieve material from a
comet intact (and we will someday) and bring it back to earth one thing to try is to distribute small pieces
(5-10$\mu$m) in appropriate water baths and follow their chemical evolution. We can expect that newer
experimental techniques will have developed to the point at which this will be possible. I do not predict
that this will make life but we can expect to find whether these processes of directed chemical evolution
occur and at what rate. In a way, this is one phase of the COSIMA project where dust from the comet
will land on a water saturated surface before being subject to mass spectroscopy.

Thought is already being directed by Krueger, Kissel and the author to simulating the porous structure of
comet dust in which laboratory organics are embedded and placing these structures in water. It is just a
matter of providing such a model to determine how effectively the chemical evolution is directed.

\acknowledgements
We would like to thank first David Block for the great effort he has put into making this event a real
success. Secondly we thank the Anglo American Chairmans Fund and the SASOL for the support 
which made the meeting and the participation of JMG
possible. One of us (C. Shen) wishes to thank the World Laboratory for a fellowship.
And finally we wish to thank Dr. Schutte and Dr. R\"{o}ttgering for advice and Dr. Li for his invaluable
assistance.

\end{article}
\end{document}